\documentclass[aip, jcp, preprint]{revtex4-1}

\usepackage{amsmath, amsfonts}
\usepackage{graphicx, url}
\usepackage{natbib}
\usepackage{multirow, rotating}
\usepackage{lineno}
\usepackage{hyperref}
\hypersetup{colorlinks = true, linktocpage = true, bookmarksopen = true, linkcolor = blue, urlcolor=blue, citecolor = blue, allcolors=blue}
\usepackage{color}
\def\cbl{\color{black}}
\def\cb{\color{black}}

\usepackage{color}
\def\cbl{\color{black}}
\def\cb{\color{black}}

\begin{document}
\title{\cbl Rapid calculation of maximum particle lifetime for diffusion in complex geometries \cb}
\author{Elliot J Carr}
\author{Matthew J Simpson}
\affiliation{School of Mathematical Sciences, Queensland University of Technology, Brisbane, Australia.}

\begin{abstract}
Diffusion of molecules within biological cells and tissues is strongly influenced by crowding. A key quantity to characterize diffusion is the \textit{particle lifetime}, which is the time taken for a diffusing particle to exit by hitting an absorbing boundary. Calculating the particle lifetime provides valuable information, for example, by allowing us to compare the timescale of diffusion and the timescale of reaction, thereby helping us to develop appropriate mathematical models.  Previous methods to quantify particle lifetimes focus on the \textit{mean particle lifetime}. Here, we take a different approach and present a simple method for calculating the \textit{maximum particle lifetime}.  This is the time after which only a small specified proportion of particles in an ensemble remain in the system. Our approach produces accurate estimates of the maximum particle lifetime, whereas the mean particle lifetime always underestimates this value compared with data from stochastic simulations. Furthermore, we find that differences between the mean and maximum particle lifetimes become increasingly important when considering diffusion hindered by obstacles.
\end{abstract}
\maketitle

\section{Introduction}
\noindent
Diffusion of molecules and cells is key to many biological processes~\cite{Murray02}.  For biological applications, it is often relevant to simulate diffusion stochastically to capture fluctuations~\cite{Lotstedt2016}.  Another important feature is to consider the role of crowding and obstacles~\cite{Cianci2016}. Intracellular and extracellular environments often contain a large volume fraction of immobile, impenetrable obstacles~\cite{Fanelli2010}.  These obstacles influence both physical~\cite{Ellery2014,Ellery2016} and chemical processes~\cite{Schnell2004,Isaacson2006}.

If we consider a stochastic model of diffusion on a domain with an absorbing boundary, a diffusing particle will be removed when it hits that boundary, giving rise to the concept of the particle lifetime~\cite{Simpson2015}, which is a special case of the first passage time~\cite{Redner01}.  When considering an ensemble of simulations, the mean of the distribution of particle lifetime, called the mean particle lifetime, is often used as a characteristic timescale of diffusion~\cite{Ellery2012b,Meinecke2017}.  While for some purposes it may be suitable to characterize a diffusive timescale by the mean of such a distribution~\cite{Berezhkovskii2011,Gordon2013,Simpson2017}, for other purposes, such as estimating the maximum time that a particle spends diffusing in a particular domain,  we are more interested in the long time behaviour of the distribution.

\section{Analysis}

\noindent
\textit{Stochastic simulations}. We simulate particle lifetime distributions using a continuous space, discrete time random walk~\cite{Codling2008}.  Time is discretized with constant time steps of duration $\tau > 0$.  In each time step a particle, at location $\mathbf{x}(t)= (x(t),y(t))$, attempts to step a distance $\delta > 0$, to $\mathbf{x}(t+\tau)= (x(t+\tau), y(t + \tau))= (x(t) + \delta \cos \theta, y(t) + \delta \sin \theta)$ with probability $\mathcal{P} \in [0,1]$.  Here, $\theta$ is sampled from a uniform distribution, $\theta \sim \mathcal{U}[0, 2\pi]$.  This discrete process corresponds to a random walk with diffusivity $D = \mathcal{P} \delta^2/(4 \tau)$.  Any potential step that would place the particle across a reflecting boundary is aborted. Simulations proceed until the particle hits an absorbing boundary, and the time taken for the particle to reach the absorbing boundary is recorded.  Repeating this procedure many times with the same choice of starting location, $\mathbf{x}_{0}=(x(0),y(0))$, enables us to construct a histogram of the particle lifetimes.\\

\noindent
\textit{Continuum description}. If $p(\mathbf{x}, t)$ is the probability of finding the particle at location $\mathbf{x}$, at time $t$, standard arguments show that $p(\mathbf{x}, t)$ evolves according to a linear diffusion equation~\cite{Codling2008},
\begin{equation}\label{eq:diffusion}
\dfrac{\partial p(\mathbf{x},t)}{\partial t} = D\nabla^{2}p(\mathbf{x},t),\quad \mathbf{x}\in\Omega,\\
\end{equation}
with $p(\mathbf{x},0) = \delta(\mathbf{x}-\mathbf{x}_{0})$; $p(\mathbf{x},t) = 0$ on absorbing boundaries, $\mathbf{x}\in\partial\Omega_{\mathrm{a}}$; and $\nabla p(\mathbf{x},t)\cdot\mathbf{n} = 0$ on reflecting boundaries, $\mathbf{x}\in\partial\Omega_{\mathrm{r}}$, where $\mathbf{n}$ is an outward facing, unit normal vector. Since we always consider problems with absorbing boundaries, the particle will always eventually leave the system and  $\displaystyle{\lim_{t \to \infty} p(\mathbf{x},t) = 0}$.   Further, as Eq (\ref{eq:diffusion}) is parabolic, $p(\mathbf{x},t)$ decays to zero exponentially fast.

The characteristic timescale associated with this process is often written in terms of the mean particle lifetime, which obeys an elliptic partial differential equation (PDE) (Supplementary Material). While it is standard to characterize the timescale of a diffusive process by simply examining a mean timescale~\cite{Berezhkovskii2011,Gordon2013,Jazaei2017}, it is well-known that diffusive processes take an infinite amount of time to proceed to completion, and that working with just the first moment can sometimes be insufficient~\cite{Carr2017}. Therefore, we consider higher moments of the particle lifetime distribution. Following similar arguments (Supplementary Material), \cbl the first $k$ raw moments are given by a family of elliptic PDEs~\cite{Redner01}, \cb
\begin{linenomath}
\begin{align}\label{eq:rawmoments}
D \nabla^2 M_{k}(\mathbf{x}) &= -kM_{k-1}(\mathbf{x}),\quad \mathbf{x}\in\Omega,
\end{align}
\end{linenomath}
for all $k\in\{1,2,3,\ldots\}$. Here, $M_{0}(\mathbf{x}) = 1$, $M_1(\mathbf{x})$ is the mean particle lifetime, and $M_{k}(\mathbf{x})$ is the $k$th raw moment of the particle lifetime distribution. To solve Eq (\ref{eq:rawmoments}), appropriate boundary conditions, for all $k\in\{1,2,3,\ldots\}$, are to set $M_{k}(\mathbf{x})=0$ along all absorbing boundaries, $\mathbf{x}\in\partial\Omega_{\mathrm{a}}$, and $\nabla M_{k}(\mathbf{x}) \cdot \mathbf{n} =0$ along all reflecting boundaries, $\mathbf{x}\in\partial\Omega_{\mathrm{r}}$.

For an arbitrary geometry, Eq (\ref{eq:rawmoments}) can be solved numerically for $M_{k}(\mathbf{x})$. To do this we use a finite volume method to discretize the governing equations over an unstructured triangular meshing of $\Omega$. \cbl To perform these calculations we use mesh generation software, GMSH~\cite{Geuzaine09}, being careful to place a node at $\mathbf{x}_{0}$. \cb The finite volume method is implemented using a vertex centered strategy with nodes located at the vertices in the mesh and control volumes constructed around each node by connecting the centroid of each triangular element to the midpoint of its edges \cite{Carr2014,Carr2016}. Linear finite element shape functions~\cite{Dhatt12} are used to approximate gradients in each element. Assembling the finite volume equations yields a linear system:
\begin{linenomath}
\begin{align}
\label{eq:linear_system}
\mathbf{A}\mathbf{M}_{k} = \mathbf{b}_{k}.
\end{align}
\end{linenomath}
The entries of the solution vector, $\mathbf{M}_{k}$, provide the value of the $k$th raw moment at each node in the mesh. A subscript $k$ is used on $\mathbf{b}_{k}$ to denote dependence on the index $k$.

Using the computed values of the raw moments at the starting location, $\mathbf{x}_{0}$, our aim is to calculate the maximum particle lifetime, or the time after which only a small, specified proportion of particles remain in the system.  Let $T$ be a continuous random variable representing the lifetime for a particle starting at $\mathbf{x}_{0}$.  Suppose $T$ has probability density function $f(t;\mathbf{x}_{0})$. The \textit{maximum particle lifetime}, $t^{\ast}$, satisfies $\mathbb{P}(T\geq t^{\ast}) = \varepsilon$, where $\varepsilon\ll 1$ is a small user-specified probability, representing the proportion of particles remaining at $t = t^{\ast}$. Equivalently, $t^{\ast}$ satisfies
\begin{linenomath}
\begin{align}
\label{eq:te}
F(t^{\ast};\mathbf{x}_{0}) = 1-\varepsilon,
\end{align}
where $F(t; \mathbf{x}_{0}) = \mathbb{P}(T\leq t) = \int_{0}^{t} f(s;\mathbf{x}_{0})\,\mathrm{d}s$ is the cumulative distribution function (CDF) of $T$. Within this framework, the $k$th moment of $T$ is:
\begin{align*}
M_{k}(\mathbf{x}_{0}) = \mathbb{E}(T^{k}) = \int_{0}^{\infty} t^{k}f(t;\mathbf{x}_{0})\,\mathrm{d}t.
\end{align*}
\end{linenomath}

Since $\varepsilon \ll 1$, it is convenient to estimate $t^{\ast}$ by first replacing $F(t^{\ast}; \mathbf{x}_{0})$ in Eq (\ref{eq:te}) with an estimate of its long time behaviour by noting that $F(t; \mathbf{x})$ can be written as,
\begin{equation}\label{eq:CDF}
F(t; \mathbf{x}_{0}) = 1 - \sum_{n=1}^{\infty}\alpha_{n}\textrm{e}^{-t\beta_{n}} \sim 1 - \alpha_{1} e^{-t\beta_{1}},
\end{equation}
as $t\rightarrow\infty$, where $\alpha_{n}$ and $\beta_{n} >0$ are constants that depend on $\mathbf{x}_{0}$ and satisfy $\beta_{1}<\beta_{2}<\hdots$ and $\alpha_{1}\neq 0$.  This form for $F(t; \mathbf{x}_{0})$ follows from $F(t;\mathbf{x}_{0}) = 1 - \int_{\Omega} p(\mathbf{x},t)\,\mathrm{d}x\,\mathrm{d}y$ and the fact that $p(\mathbf{x}, t)$ decays exponentially fast as $t \to \infty$.

Under these assumptions, the constants that appear in the dominant term of the summation in Eq (\ref{eq:CDF}) can be identified in terms of the moments\cite{Carr2017}:
\begin{linenomath}
\begin{align}
\label{eq:alpha_beta}
\alpha_{1} &\sim \frac{M_{k}(\mathbf{x}_{0})}{k!}\left(\frac{k M_{k-1}(\mathbf{x}_{0})}{M_{k}(\mathbf{x}_{0})}\right)^{k}\, \text{and} \,   \notag \\
\beta_{1} &\sim \frac{k M_{k-1}(\mathbf{x}_{0})}{M_{k}(\mathbf{x}_{0})},
\end{align}
\end{linenomath}
as $k\rightarrow\infty$. \cbl Writing the constants $\alpha_1$ and $\beta_1$ in terms of the moments is possible because of the long time exponential behaviour of $F(t;\mathbf{x}_{0})$.  This approach may not be applicable if we were considering some other form of long time behaviour~\cite{Gumbel78}. \cb Substituting Eq (\ref{eq:CDF}) into Eq (\ref{eq:te}), solving for $t^{\ast}$ and inserting the expressions (\ref{eq:alpha_beta}) yields the following asymptotic estimate of the maximum particle lifetime:
\begin{linenomath}
\begin{align}
\label{eq:exittime}
t^{\ast}(\mathbf{x}_{0}) \sim \frac{M_{k}(\mathbf{x}_{0})}{kM_{k-1}(\mathbf{x}_{0})}\ln\left[\frac{M_{k}(\mathbf{x}_{0})}{k!\,\varepsilon}\left(\frac{kM_{k-1}(\mathbf{x}_{0})}{M_{k}(\mathbf{x}_{0})}\right)^{k}\right],
\end{align}
\end{linenomath}
as $k\rightarrow\infty$ and $\varepsilon\rightarrow 0$.

In summary, the procedure for calculating $t^{\ast}$ for a chosen value of $\varepsilon$ involves looping over the index $k$ and at every loop iteration: (i) solving Eq (\ref{eq:rawmoments}); (ii) applying Eq (\ref{eq:exittime}) to calculate $t^{\ast}$ using $M_{k-1}(\textbf{x})$ and $M_{k}(\textbf{x})$; and (iii) terminating the iterations once $t^{\ast}$ converges. We find that using the first 5--10 moments gives useful results.

\section{Results}

\begin{figure*}[p]
\centering
\vspace*{-2.0cm}\hspace*{1.5cm}\includegraphics[width=0.90\textwidth]{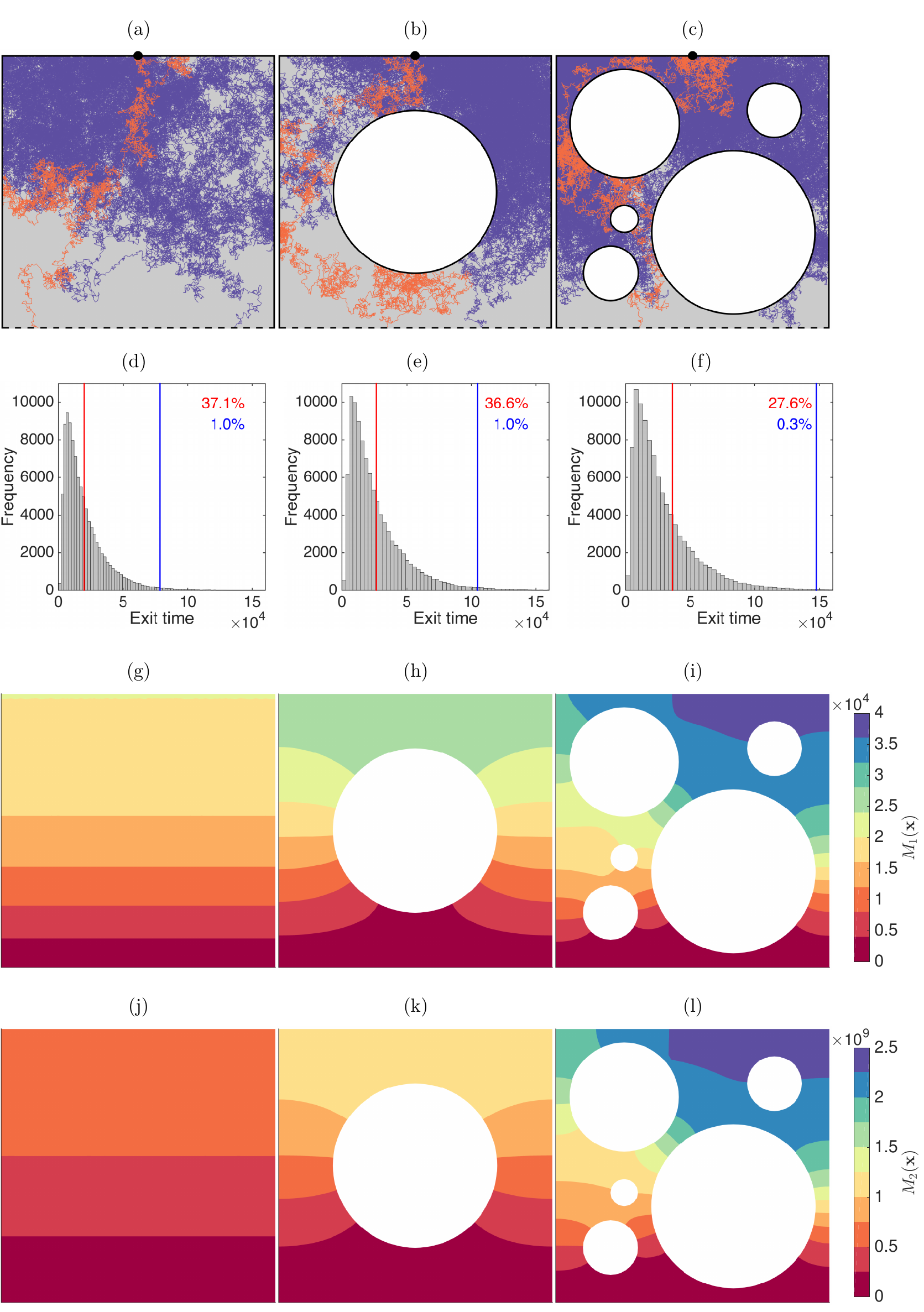}
\caption{(a)--(c) Geometry and stochastic simulations for three test cases. (d)--(f) Histogram of the particle lifetimes using 100,000 simulations with $\mathcal{P}=1$, $\delta = 0.01$, $\tau = 1$ and $\mathbf{x}_{0}=(0.5,1)$ ($\bullet$).  Vertical lines depict the mean particle lifetime, $M_{1}(\mathbf{x}_{0})$ (red), and the maximum particle lifetime, $t^{\ast}(\mathbf{x}_{0})$ (blue), computed using $k = 10$ and $\varepsilon=0.01 \cbl= 1\%\cb$. (g)--(i) and (j)--(l) Contour plots of $M_{1}(\mathbf{x})$ and $M_{2}(\mathbf{x})$, respectively. The meshes used to solve Eq (\ref{eq:rawmoments}) were generated in GMSH by prescribing a mesh element size of 0.02 producing the following numbers of nodes: 3436 (no obstacles), 2518 (one obstacle) and 1982 (several obstacles).}
\label{fig:1}
\end{figure*}

We now present some results to illustrate the benefit of our method.  We first consider the canonical problem of a random walk in a unit square with $\mathbf{x}_0=(0.5,1.0)$. The domain and sample trajectories are shown in Figure \ref{fig:1}a.   In this case we have an absorbing boundary along the lower horizontal boundary (dashed line), and reflecting boundary conditions elsewhere (solid line).  Performing 100,000 identically prepared realizations of the stochastic model allows us to construct a histogram of the particle lifetime information, shown in Figure \ref{fig:1}d.

Stochastic results are compared with our analysis by solving Eq (\ref{eq:rawmoments}) for $M_{k}(\mathbf{x})$ sequentially for $k=1,2,3,\ldots,$. \cbl For this simple geometry we can solve for $M_{k}(\mathbf{x})$ exactly, giving $M_1(\textbf{x}_0)= 2\times10^{4}$, \cb and we have a good match between the moments given by Eq (\ref{eq:rawmoments}), and by estimating the moments directly from the stochastic data.  It is useful to note that the mean particle lifetime is a very poor estimate of the maximum particle lifetime because when $t=M_1(\textbf{x}_{0})$, there are still approximately 37\% of particles in the ensemble present in the domain (Figure \ref{fig:1}d).  In contrast, using Eq (\ref{eq:exittime}) with $\varepsilon=0.01$, we find $t^{\ast}\approx 7.86\times10^4$ and this corresponds to the time when just 1\% of particles remain (Figure \ref{fig:1}d).  The solutions for the first two raw moments, $M_1(\textbf{x})$ and $M_2(\textbf{x})$ are given in Figures \ref{fig:1}g and \ref{fig:1}j, respectively.  Due to the absence of obstacles and the choice of boundary conditions, the moments are independent of horizontal position.

We repeat the simulations and calculations for more biologically relevant domains that contain obstacles.  As before, we consider a random walk on a unit square except now we consider configurations in Figure \ref{fig:1}b--c with various arrangements of obstacles. The geometry in Figure \ref{fig:1}b contains one circular obstacle of radius 0.3 centered at $(0.5,0.5)$. The geometry in Figure \ref{fig:1}c contains five circular obstacles centered at $(0.65,0.35), (0.8,0.8), (0.2,0.2), (0.25,0.75)$ and $(0.25,0.4)$ with radii 0.3, 0.1, 0.1, 0.2 and 0.05, respectively. For both configurations, the lower horizontal boundary of the domain is an absorbing boundary (dashed line) while all other boundaries are reflecting (solid line). In Figures \ref{fig:1}b--c, two example particle paths are shown corresponding to a small (orange) and large (purple) particle lifetime, respectively. For each domain we perform 100,000 identically prepared realizations of the stochastic model and construct histograms of the particle lifetime data in Figure \ref{fig:1}e--f, respectively. From the solutions, plotted in Figures \ref{fig:1}h--i, we have $M_1(\textbf{x}_{0}) \approx 2.64\times10^{4}$  and $M_{1}(\textbf{x}_{0}) \approx 3.62\times 10^{4}$ for the problems with one and several obstacles, respectively. \cbl Considering our calculations $t^{\ast}\approx 1.05\times10^5$ and $t^{\ast}\approx 1.48\times10^5$, for the problem with one and several obstacles, respectively, the mean particle lifetime is comparatively insensitive (in an absolute sense) to the presence and arrangement of obstacles.  This shows that standard mean particle lifetime data may not always provide a useful way to distinguish between the presence or absence of obstacles. \cb This is particularly evident in Figures \ref{fig:1}d--f, where the difference between $M_1(\textbf{x}_0)$ (red vertical line) and $t^{\ast}$  (blue vertical line) increases as the number of obstacles is increased. For the domain with one obstacle, at $t=M_1(\textbf{x}_{0})$, we have approximately 37\% of particles from the ensemble still present in the domain, whereas at $t=t^{\ast}$ we have just 1\% of particles remaining. Similar trends are observed for the problem with several obstacles.

\cbl Additional numerical results for the test case with one obstacle are given in the Supplementary Material (Section II). These results demonstrate the rapid convergence exhibited by the asymptotic estimate, Eq (\ref{eq:exittime}), with only the first six moments required to produce a converged estimate of $t^{\ast}$, accurate to within four significant figures (Table I, Supplementary Material). These results also highlight the accuracy of the asymptotic estimate, Eq (\ref{eq:exittime}), in the range $\varepsilon = 10^{-4}$--$10^{-1}$ with the proportion of particles remaining at $t = t^{\ast}$ a close match with the specified value of $\varepsilon$ in each case. Note that $\varepsilon$ has no effect on the computational efficiency of our approach as it is simply an input parameter in Eq (\ref{eq:exittime}). The additional computational cost of incrementing the number of moments $k$ by one, which requires the solution of one additional linear system (\ref{eq:linear_system}), is minimal (for our choice of mesh resolution) so using the converged estimate of $t^{\ast}$ is practicable. In comparison, using a random walk simulation with fewer realizations to estimate the maximum particle lifetime takes longer (approximately 10 seconds versus less than one second for the asymptotic estimate, Eq~(\ref{eq:exittime})) and produces less accurate estimates of $t^{\ast}$ (Table I, Supplementary Material).
\cb

Overall, (i) our calculation of $t^{\ast}$ is a more accurate estimate of the maximum particle lifetime than standard mean particle lifetime calculations; (ii) our approach avoids the need for performing any stochastic simulations; and (iii) our approach allows $t^{\ast}$ to be calculated for \textit{any} starting location as the raw moments, $M_{k}{\textbf(\mathbf{x})}$ for $k=1,2,3,\ldots$, are known at all locations in the domain. Together, this means that our technique is accurate, efficient and avoids any stochastic simulations. 

\section{Discussion}
Here we describe a fast, new and simple-to-implement approach to calculate the maximum particle lifetime for diffusion. We also show that there can be major differences between the mean timescale and the maximum timescale, and that these differences are sensitive to the presence and arrangement of obstacles in the domain. Our approach is more accurate than using the mean time scale.  For example, in our calculations at $t=M_1(\textbf{x}_0)$ with $\varepsilon = 1\%$, there are still approximately 30--40\% of particles remaining in the domain.  In contrast, at $t=t^{\ast}$, there are approximately 1\% of particles remaining in the domain.  \cbl Furthermore, our approach is computationally efficient. For example, carrying out the 100,000 random walk simulations and generating the histograms in Figures \ref{fig:1}d--f requires approximately 1--3 minutes of computation time on a single desktop machine, whereas solving Eq (\ref{eq:rawmoments}) for $k = 1,2,\hdots,10$ and applying Eq (\ref{eq:exittime}) takes less than one second across all three test cases.\cb

\cbl Estimates of $t^{\ast}$ could be obtained by solving Eq (\ref{eq:diffusion}) numerically and calculating the duration of time required for the survival probability, $\int_{\Omega} p(\mathbf{x},t)\,\mathrm{d}x\,\mathrm{d}y$, to decay to $\varepsilon$.   We do not recommend this approach because our moment-based method is far more efficient.  To demonstrate this difference in efficiency we note that the computational cost to solve Eq (\ref{eq:diffusion}) using standard standard spatial and implicit temporal discretization methods can be quantified in terms of the number of times that the solution of a linear system, of size $n \times n$, where $n$ denotes the number of nodes in the spatial discretization, is computed. Our approach requires the solution of $m+1$ linear systems, of size $n\times n$, for $k = m$.  Calculating the transient solution of Eq (\ref{eq:diffusion}) requires the solution of a linear system, of size $n\times n$, at each time step if the same spatial mesh is used in both calculations.  We obtain  accurate estimates  of $t^{\ast}$ using $m = 5 - 10$.  This estimate requires the solution of no more than eleven linear systems. In contrast, the number of time steps required to solve Eq (\ref{eq:diffusion}) for sufficiently large $t$ so that the survival probability decays sufficiently could be many orders of magnitude greater.   \cb

\cbl
Our approach to calculating $t^\ast$ relies on the fact that the CDF, Eq (\ref{eq:CDF}), can be expressed in terms of a sum of exponential functions, and that the large time behaviour of the CDF is approximately exponential.  If the large time behaviour of the CDF had some different asymptotic form, such as in the case of various extreme distributions~\cite{Gumbel78}, our approach to express the maximum exit time as the ratio of consecutive moments would not hold. However, since we are dealing with diffusion, for which we know that the solution of Eq (\ref{eq:diffusion}) always decays exponentially with time, the long time asymptotics of the CDF is always exponential, and our approach is always valid.  However, it is worthwhile to note that if we consider some a process with different long time asymptotic behaviour, such as a slowly decaying power law~\cite{Gumbel78}, a different approach would be required. \cb

\cbl
In this work, our stochastic results use a fixed step length, $\delta = 0.01$. For larger values of $\delta$ we expect the results to be less accurate as Eq (\ref{eq:diffusion})--(\ref{eq:rawmoments}) are relevant in the continuum limit as $\delta \to 0$. However, for smaller values of $\delta$ the computational advantages of our approach will be even more pronounced as each individual random walk simulations will require more steps to exit the system.
\cb

There are many ways that our calculation can be extended.  The example calculations presented here deal with just a sample of problems, and there are many other possible configurations to be explored.  Our preliminary calculations (not shown) suggest that our approach is valid for other domain geometries, obstacle geometries and obstacle densities.  Furthermore, it is possible to generalize our approach to deal with three-dimensional diffusion.

\section{Acknowledgements}
This work is supported by the Australian Research Council (DE150101137, DP170100474), \cbl and we thank two referees for helpful comments. \cb


\end{document}